\documentclass[12pt,a4paper,final]{iopart}
\usepackage{iopams}  
\usepackage{graphicx}
\usepackage[breaklinks=true,colorlinks=true,linkcolor=blue,urlcolor=blue,citecolor=blue]{hyperref}

\newcommand{\vc}{\mathbf}
\usepackage{graphicx}
\usepackage{color}

\begin{document}


\title[Instabilities in the plasma boundary layer]{Influence of Ion Streaming Instabilities on Transport Near Plasma Boundaries}


\author{Scott D.\ Baalrud}
\address{Department of Physics and Astronomy, University of Iowa, Iowa City, Iowa 52242, USA}



\date{\today}

\begin{abstract}

Plasma boundary layers are susceptible to electrostatic instabilities driven by ion flows in presheaths and, when present, these instabilities can influence transport. In plasmas with a single species of positive ion, ion-acoustic instabilities are expected under conditions of low pressure and large electron-to-ion temperature ratio ($T_e/T_i \gg 1$). In plasmas with two species of positive ions, ion-ion two-stream instabilities can also be excited. The stability phase-space is characterized using the Penrose criterion and approximate linear dispersion relations. Predictions for how these instabilities affect ion and electron transport in presheaths, including rapid thermalization due to instability-enhanced collisions and an instability-enhanced ion-ion friction force, are also briefly reviewed. Recent experimental tests of these predictions are discussed along with research needs required for further validation. The calculated stability boundaries provide a guide to determine the experimental conditions at which these effects can be expected.

\end{abstract}



\vspace{2pc}
\submitto{\PSST}


\maketitle



\section{Introduction\label{sec:intro}}

Plasma boundary layers are often thought to be laminar regions in which ions are accelerated through a presheath to a speed exceeding the ion sound speed $c_s \equiv \sqrt{k_B T_e/m_i}$ at the sheath edge~\cite{bohm:49,lieb:05,robe:13}. It has recently been suggested that under certain discharge conditions the ion flow may not be laminar, but instead excite electrostatic instabilities in the presheath. In particular, ion-acoustic instabilities are expected under conditions of low neutral pressure and large electron-to-ion temperature ratio ($T_e/T_i \gg 1$)~\cite{baal:09a,baal:11b}. If two different species of positive ions are present, ion-ion two-stream instabilities can also be expected under similar conditions~\cite{hers:05,baal:09,baal:11,baal:15,yip:10,hers:11}. This work provides a characterization of the stability boundaries for each of these, which can be used as a practical guide to determine the experimental conditions at which the instabilities are expected to arise in presheaths. 

These flow-driven instabilities have been predicted to affect transport in various ways.  At common low temperature plasma conditions, ionization and charge exchange collisions typically cause a slow tail to form in the ion velocity distribution function (IVDF) as ions flow through the presheath~\cite{sher:01,clai:06}.  For single ion species plasmas, the ion-acoustic instability has been predicted to increase the ion-ion collision rate through wave-particle scattering, causing the IVDF to thermalize as it nears the sheath edge~\cite{baal:11b}. The ion-acoustic instabilities have similarly been proposed to enhance electron scattering, providing a possible explanation for Langmuir's paradox~\cite{baal:09a}. In ionic mixtures, ion-ion two-stream instabilities have been predicted to rapidly enhance the ion-ion friction force, preventing the differential ion flow speed ($\Delta V = V_1 -V_2$) from significantly exceeding the threshold value for instability onset $(\Delta V_c$)~\cite{baal:09,baal:11,baal:15}. This condition, along with the two-species Bohm criterion, was proposed to determine the speed of each ion species at the sheath edge. Aspects of each of these predictions have since been tested experimentally~\cite{yip:10,hers:11,yip:15}. This paper provides a review of the current status of this validation effort, along with an analysis of instability threshold conditions that will aid in the design and analysis of future experiments. 

The basic physics of instabilities affecting transport near boundaries may have important consequences in various applications. For instance, the form of the IVDF at the sheath edge is important in materials processing because it influences the energy of ions striking a material surface. It can also influence global plasma models for mixtures because the predicted instability-enhanced friction between ions can influence the relative fluxes of each species exiting a plasma~\cite{kim:15}. This in turn influences the steady-state concentrations of each species in the bulk plasma. 

In addition to the presheath flow driven instabilities discussed here, other instances of similar electrostatic instabilities have been found in the plasma boundary layer.   For instance, when surfaces emit secondary electrons they are accelerated through the ion sheath and can generate instabilities~\cite{sydo:07,camp:12}. Sheath-induced instabilities can be excited in $\vc{E} \times \vc{B}$ discharges~\cite{smol:13}, and finite length and boundary effects of the ion-acoustic instabilities can also be significant~\cite{raps:14,kosh:15}. It has also recently been shown that electron sheaths have a presheath that accelerates electrons to the thermal speed at the sheath edge~\cite{yee:15}. This strong electron flow has been observed in 2D particle-in-cell (PIC) simulations to drive large-amplitude ion-acoustic instabilities in the electron presheath. These effects are similar to the presheath driven instabilities discussed here. 

This paper is organized as follows. Section~\ref{sec:1ion} discusses the stability boundaries for ion-acoustic instabilities in the presheath of plasma with one ion species. A summary is provided of recent experiments testing the influence of these instabilities on transport. Properties of the instabilities that may be useful for the design and interpretation of future experimental tests are provided.  Section~\ref{sec:2ions} discusses the stability of the presheath in binary ionic mixtures. Here, the focus is on ion-ion two-stream instabilities. Recent experimental and PIC simulation tests of the instability-enhanced friction concept are summarized, and properties of the predicted instabilities to aid future tests are provided. 


%
%

\section{One ion species\label{sec:1ion}} 

\subsection{Linear ion-acoustic instabilities} 

\subsubsection{Threshold conditions} 

The Penrose criterion~\cite{penr:60} provides a convenient way to calculate the exact solution of the stability boundaries for linear electrostatic instabilities in the absence of collisions. The dispersion relation is obtained from the roots of the dielectric response function 
\begin{equation}
\hat{\varepsilon} (\vc{k}, \omega) = 1 + \sum_s \frac{4\pi q_s^2}{k^2 m_s} \int d^3v \frac{\vc{k} \cdot \partial f_{s,o}(\vc{v})/\partial \vc{v}}{\omega - \vc{k} \cdot \vc{v}} , \label{eq:ephat}
\end{equation} 
where $f_{s,o}$ is the lowest order velocity distribution function of species $s$. The Penrose criterion states that a necessary and sufficient condition for instability is given by
\begin{equation}
P(F) \equiv \int_{-\infty}^\infty du \frac{F(u) - F(u_o)}{(u-u_o)^2} > 0 \label{eq:penrose}
\end{equation}
where 
\begin{equation}
F(u) \equiv \sum_s \frac{4\pi q_s^2}{m_s} \int d^3v f_{o,s} (\vc{v}) \delta (u - \hat{\vc{k}} \cdot \vc{v}) ,
\end{equation}
is the total distribution function projected along the direction $\hat{\vc{k}} \equiv \vc{k}/k$. For a double peaked distribution, as we are considering here, $u_1$ and $u_2$ are the $u$ locations of the two peaks, and $u_o$ is the location of the local minimum of $F(u)$ in the region $u_1 < u_o < u_2$. The Penrose criterion can also be used to determine the range of unstable wave numbers~\cite{davi:83}. This is given by $k_{\min}^2 < k^2 < k_{\max}^2$ where $k_{\max}^2 = P(\hat{F})$ and $k_{\min} = \textrm{max} \lbrace 0, \hat{k} \rbrace$ where 
\begin{equation}
\hat{k}^2 \equiv \int_{-\infty}^\infty du \frac{F(u) - F(u_2)}{(u-u_2)^2} . \label{eq:khat}
\end{equation}


To quantify the stability boundaries, consider Maxwellian ion and electron distributions functions with a differential flow $f_{e,o} = n_e \exp (-v^2/v_{Te}^2)/(\pi^{3/2} v_{Te}^3)$ and $f_{i,o} = n_i \exp [-(\vc{v} - \vc{V}_i)^2/v_{Ti}^2]/(\pi^{3/2} v_{Ti}^3)$. For these model distributions 
\begin{equation}
F(u) = \frac{\omega_{pi}^2}{\sqrt{\pi} v_{Ti}} \exp \biggl[- \frac{(u-V_d)^2}{v_{Ti}^2} \biggr] + \frac{\omega_{pe}^2}{\sqrt{\pi} v_{Te}} \exp \biggl( - \frac{u^2}{v_{Te}^2} \biggr) \label{eq:fuia}
\end{equation}
where $\omega_{ps} = \sqrt{4\pi q_s^2 n_s/m_s}$ is the plasma frequency of species $s$, $v_{Ts} \equiv \sqrt{2 k_B T_s/m_s}$ is the thermal speed of species $s$ and $V_d = \vc{V}_i \cdot \hat{k}$ is the differential drift speed between electron and ion fluids. To evaluate equations (\ref{eq:penrose}) and (\ref{eq:khat}), $u_o, u_1$ and $u_2$ were computed from $dF/du = 0$. 

\begin{figure}
\begin{center}
\includegraphics[width=7cm]{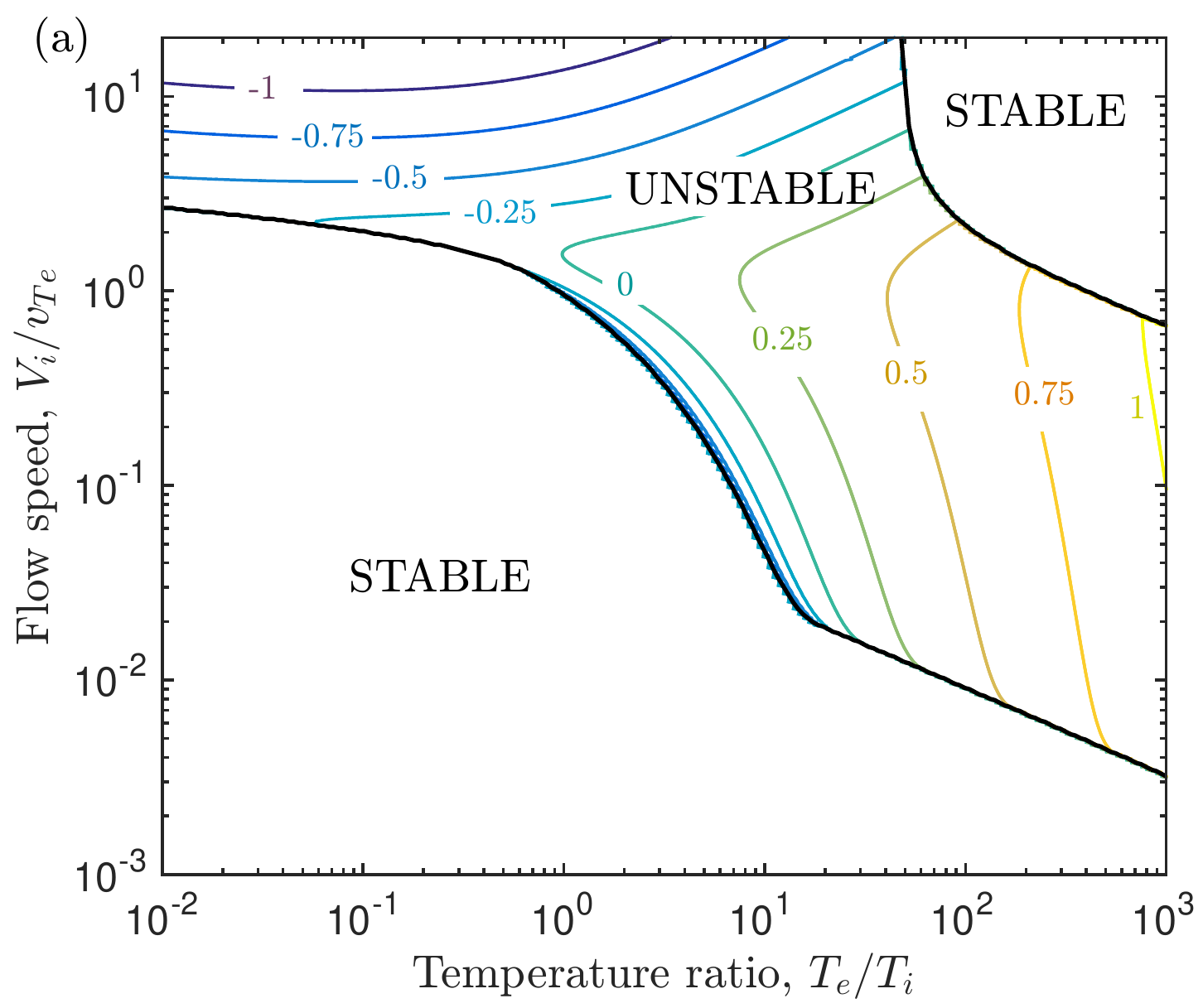}\\
\includegraphics[width=7cm]{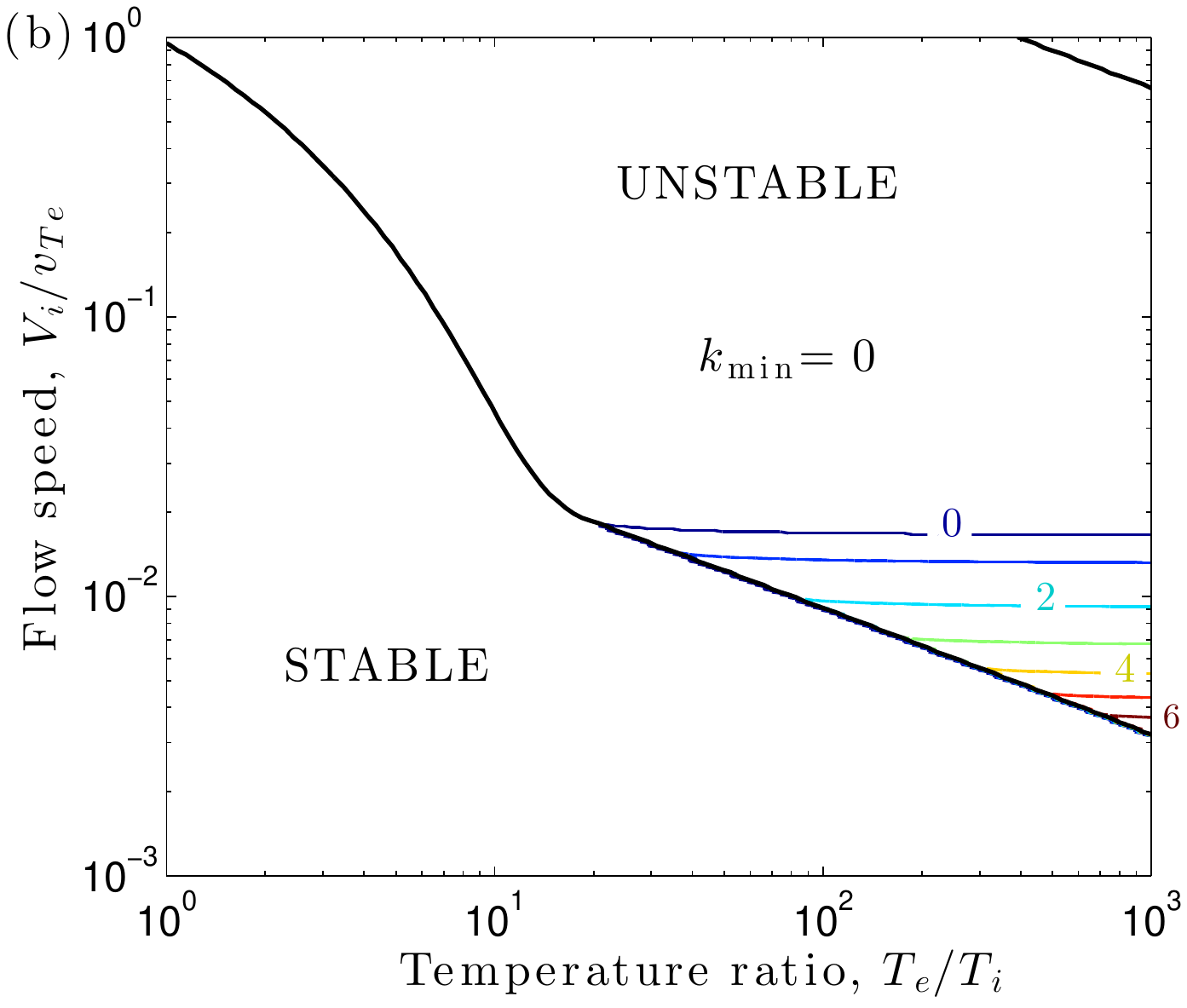}
\caption{Ion-acoustic instabilities boundaries for a plasma with H$^+$ ions. (a) Contours show lines of constant $\textrm{log}_{10} (k_{\max}\lambda_{De})$ calculated from equation~(\ref{eq:penrose}). (b) Contours show lines of constant $k_{\min} \lambda_{De}$ calculated from equation~(\ref{eq:khat}) for H$^+$.}
\label{fg:ia_vt}
\end{center}
\end{figure}

Figure~\ref{fg:ia_vt} shows the solution of equation~(\ref{eq:penrose}) for the distribution in equation (\ref{eq:fuia}) with H$^+$ ions over a broad range of temperature ratio and flow speed. Here, the flow speed is in units of $v_{Te}$. This figure provides the stability diagram required to determine if ion-acoustic instabilities can be expected under different experimental conditions. It is, however, limited to low pressure since it does not account for collisions. The contours in the unstable region of figure~\ref{fg:ia_vt}a show the $k_{\max}$ values computed from equation~(\ref{eq:penrose}) in units of electron Deybe length $\lambda_{De}$. Correspondingly, the contours in the unstable region of figure~\ref{fg:ia_vt}b show the $k_{\min}$ values computed from equation~(\ref{eq:khat}). Together, these provide the range of wavelengths that are predicted to be excited over this range of experimental conditions.  

\begin{figure}
\begin{center}
\includegraphics[width=7cm]{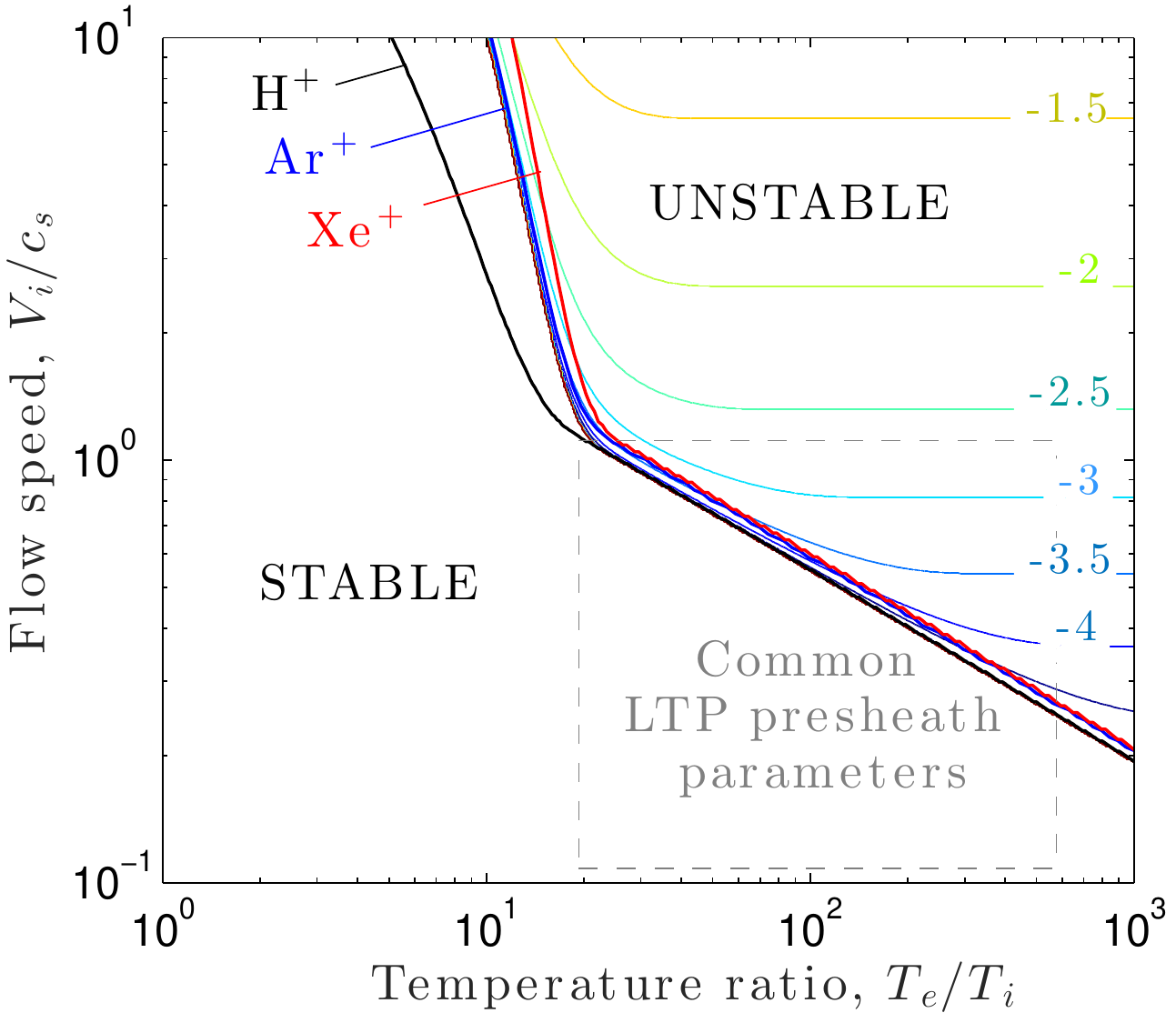}
\caption{Ion-acoustic instability boundaries for plasma with H$^+$ (black), Ar$^+$ (blue) and Xe$^+$ (red) ions calculated from equation~(\ref{eq:penrose}). Numbered contours show the maximum growth rate $\textrm{log}_{10}(\gamma_{\max}/\omega_{pi})$  for the H$^+$ plasma computed from the approximate dispersion relation in equation~(\ref{eq:gama}). }
\label{fg:ia_cs}
\end{center}
\end{figure}

Figure~\ref{fg:ia_cs} shows stability boundaries focusing on the range of parameters of interest to presheaths in low-temperature plasmas. Curves are shown for plasma with either H$^+$, Ar$^+$ or Xe$^+$ ions. Here, the flow speed is in units of $c_s$. The parameter regime common to presheaths in low temperature plasmas is indicated with a dashed line. This shows that when the temperature ratio is high $T_e/T_i \gtrsim 20$, ion-acoustic instabilities can often be expected near the sheath edge. At higher temperature ratios, a broader portion of the presheath becomes susceptible to the instability. 

\begin{figure}
\begin{center}
\includegraphics[width=7cm]{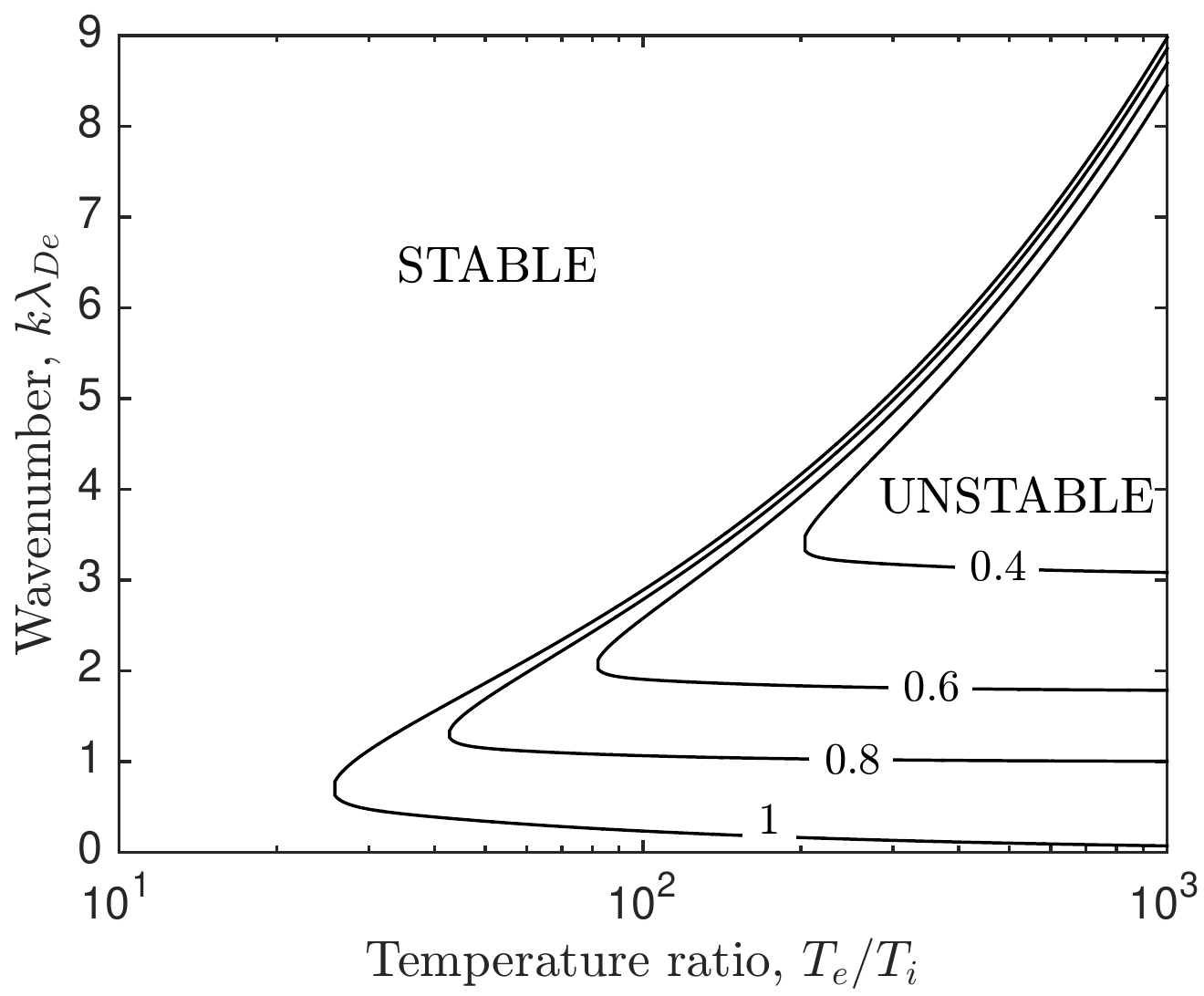}
\caption{Wavnumber boundaries for the ion-acoustic instability in an H$^+$ plasma calculated from equations~(\ref{eq:penrose}) and (\ref{eq:khat}) for four values of the ion drift speed: $V_i/c_s$ = 0.4, 0.6, 0.8 and 1.}
\label{fg:ia_k}
\end{center}
\end{figure}

Figure~\ref{fg:ia_k} shows the range of unstable wavenumbers for parameters relevant to the presheath of a low-temperature H$^+$ plasma. These are computed from equation~(\ref{eq:penrose}) (top of the curves) and (\ref{eq:khat}) (bottom of the curves). Contours are shown for four values of the ion drift speed, corresponding to the conditions at different locations in the presheath. The figure indicates that a broader spectrum of unstable waves is expected as $T_e/T_i$ increases, and that the wave spectrum broadens as the flow speed increases toward the sheath edge. For most of the presheath, the range of unstable wavenumbers is larger than $\lambda_{De}^{-1}$, indicating short-wavelength instabilities. These contours provide the wavenumber spectrum for which future experiments, or particle simulations, might search for the ion-acoustic instability in the presheath of low pressure plasmas. 

\subsubsection{Growth rates} 

The Penrose criterion provides exact solutions for the threshold conditions of the collisionless linear dispersion relation, but it provides no information about the growth rate. Here, we evaluate a common approximate expression for the growth rate of the ion-acoustic instability for the conditions of a low-temperature plasma presheath. The dispersion relation can be computed from the roots of the linear dielectric function 
\begin{equation}
\hat{\varepsilon}(\vc{k}, \omega) = 1 - \frac{\omega_{pi}^2}{k^2 v_{Ti}^2} Z^\prime \biggl( \frac{\omega - \vc{k} \cdot \vc{V}_i}{kv_{Ti}} \biggr) - \frac{\omega_{pe}^2}{k^2 v_{Te}^2} Z^\prime \biggl( \frac{\omega}{kv_{Te}} \biggr) \label{eq:ephat_ia}
\end{equation}
in which $Z^\prime(w)$ is the derivative of the plasma dispersion function. Searching for ion-acoustic instabilities in the $T_e/T_i \gg 1$ regime, we take the large argument expansion for the ion term $Z^\prime(w\gg 1) \simeq w^{-2} - 2i w \sqrt{\pi} \exp (-w^2)$ and the small argument expansion of the electron term $Z^\prime (w \ll 1) \simeq - 2 - 2iw\sqrt{\pi}$. Inserting these into equation~(\ref{eq:ephat_ia}) and solving for the dispersion relation from $\hat{\varepsilon} = 0$, assuming that $\Im \lbrace \omega \rbrace \ll \Re \lbrace \omega \rbrace$, one arrives at a common approximation for the dispersion relation: $\omega = \omega_R + i \gamma$. Here, the real part of the wave frequency is 
\begin{equation}
\omega_r = \vc{k} \cdot \vc{V}_i - \frac{kc_s}{\sqrt{1 + k^2\lambda_{De}^2}}
\end{equation}
and the growth rate is
\begin{eqnarray}
\gamma &=& - \frac{k c_s \sqrt{\pi/8}}{(1+k^2\lambda_{De}^2)^{2}} \biggl\lbrace \biggl( \frac{T_e}{T_i} \biggr)^{3/2} \exp \biggl( - \frac{T_e/T_i}{2 (1 + k^2 \lambda_{De}^2)} \biggr) \label{eq:gama} \\ \nonumber
& + & \sqrt{ \frac{m_e}{m_i}} \biggl( 1 - \frac{V_i}{c_s} \sqrt{1 + k^2 \lambda_{De}^2} \biggr) \biggr\rbrace  .
\end{eqnarray}

\begin{figure}
\begin{center}
\includegraphics[width=7cm]{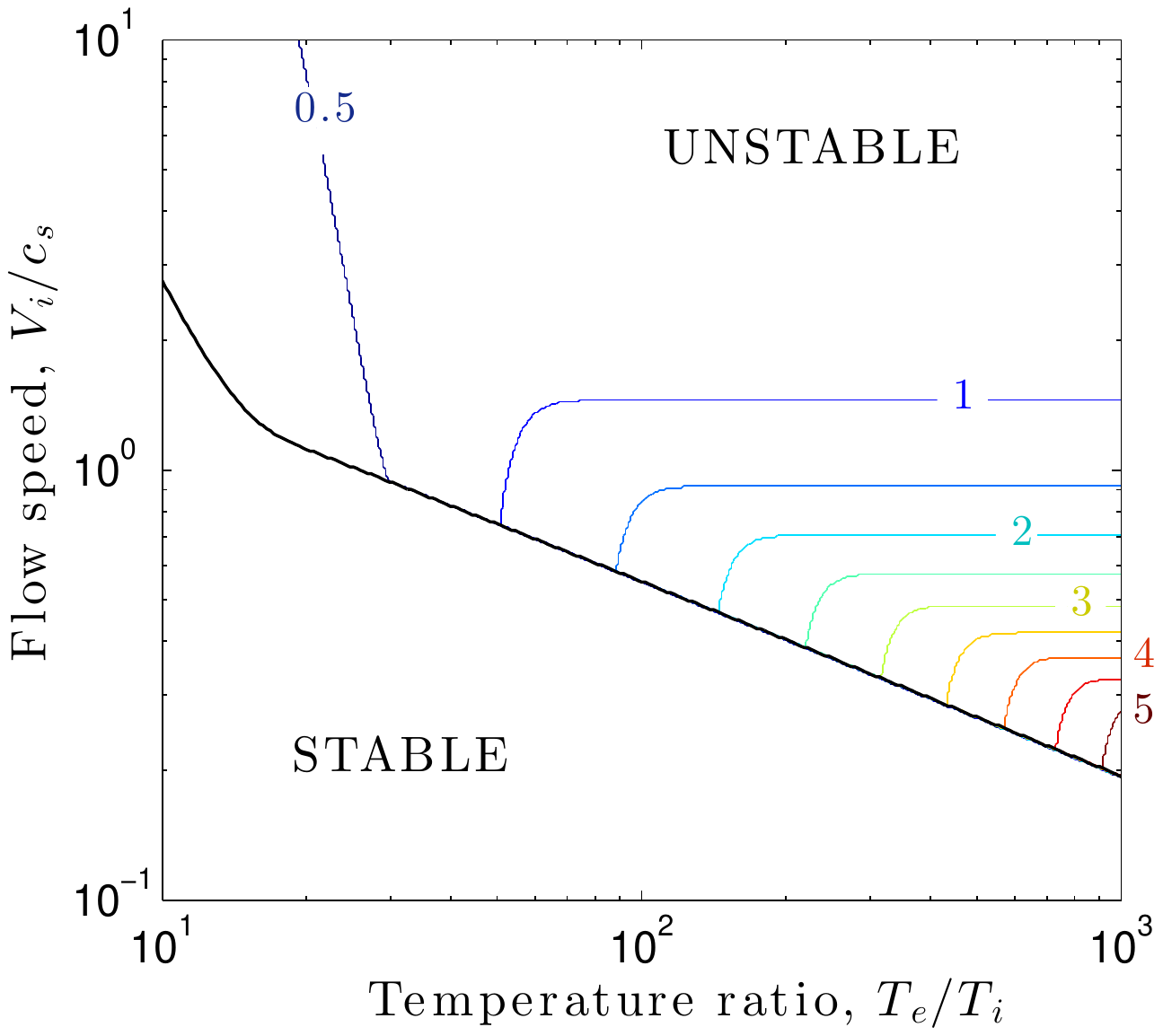}
\caption{Contours of the wavenumber (in units of $\lambda_{De}^{-1}$) corresponding to the maximum growth rate of ion-acoustic instabilities. This was computed from the approximate dispersion relation in equation~(\ref{eq:gama}).}
\label{fg:ia_kmax}
\end{center}
\end{figure}

Figure~\ref{fg:ia_cs} shows contours of the maximum growth rate for a H$^+$ plasma computed from equation~(\ref{eq:gama}). The peak growth rate in the parameter regime relevant to presheaths is much smaller than the ion plasma frequency ($\approx 10^{-3} \omega_{pi}$ near the sheath edge). The figure shows that the approximate dispersion relation accurately captures the instability boundary when $T_e/T_i \gtrsim 20$. Below this value, the approximate dispersion relation should not be considered reliable. 

In addition to the range of unstable wavenumbers, as shown in figure~\ref{fg:ia_k}, the wavenumber corresponding to the maximum growth rate is also of interest experimentally because this represents the dominant mode.  Figure~\ref{fg:ia_kmax} shows contours of the wavenumber corresponding to $\gamma_{\max}$ over a range of parameters relevant to the presheath. This was computed from the approximate dispersion relation in equation~(\ref{eq:gama}). The dominant wavelength is found to be a few Debye lengths ($\lambda = 2\pi/k$) for the unstable region of a presheath. It is also interesting to note that the group velocity of the unstable waves, $v_g =d \omega_r/dk = V_i - c_s/(1+k^2\lambda_{De}^2)^{3/2}$, is slower than the ion drift. For the values shown in figure~\ref{fg:ia_k}, the group speed of the fastest growing mode typically varies from (0.4-0.8$c_s$) through the presheath. 

\subsubsection{Neutral damping} 

The ionization fraction in low-temperature plasmas is often on the order of a few percent and ion-neutral collisions can be an important process affecting transport. In fact, the presheath length scale is often determined by the ion-neutral collision mean free path~\cite{oksu:02}. With regard to ion-acoustic instabilities, which are longitudinal waves in the ion fluid, these collisions cause an energy sink that damps the waves. This reduces the instability growth rate and, when the collision rate is sufficiently high, can completely suppress the instabilities. Thus, there is also a neutral pressure threshold to consider for ion-acoustic instabilities to be expected in a presheath. This is an important threshold to consider as experiments are found over a broad range of neutral pressures, spanning both sides of this neutral pressure threshold. 

The Penrose criterion is limited to a collisionless approximation. The following estimates for the neutral pressure threshold are based on the approximate dispersion relation from equation~(\ref{eq:gama}). To estimate wave damping by ion-neutral collisions, a simple BGK-type collision model $\nu (f_i-f_o)$ is applied to the kinetic theory, where $\nu$ is the ion-neutral collision frequency~\cite{bhat:54}. In general this is velocity dependent, but in the region of a presheath susceptible to ion-acoustic instability the ion drift is much larger than the either the ion thermal speed or the thermal speed of neutrals $V_i/v_{Ti} \simeq V_i/v_{Tn} \simeq T_e/T_i \gg 1$ (we are considering only large temperature ratio here). With this consideration, the ion-neutral collision frequency is approximated using the velocity value associated with the drift. This is determined from the dominant ion-neutral collision cross section $\nu = n_g \langle \sigma v \rangle \simeq n_g \sigma(V_i) V_i$, where $n_g$ is the neutral gas density, $\sigma$ is the ion-neutral collision cross section and $\langle \ldots \rangle$ denotes a velocity average. 

\begin{figure}
\begin{center}
\includegraphics[width=7cm]{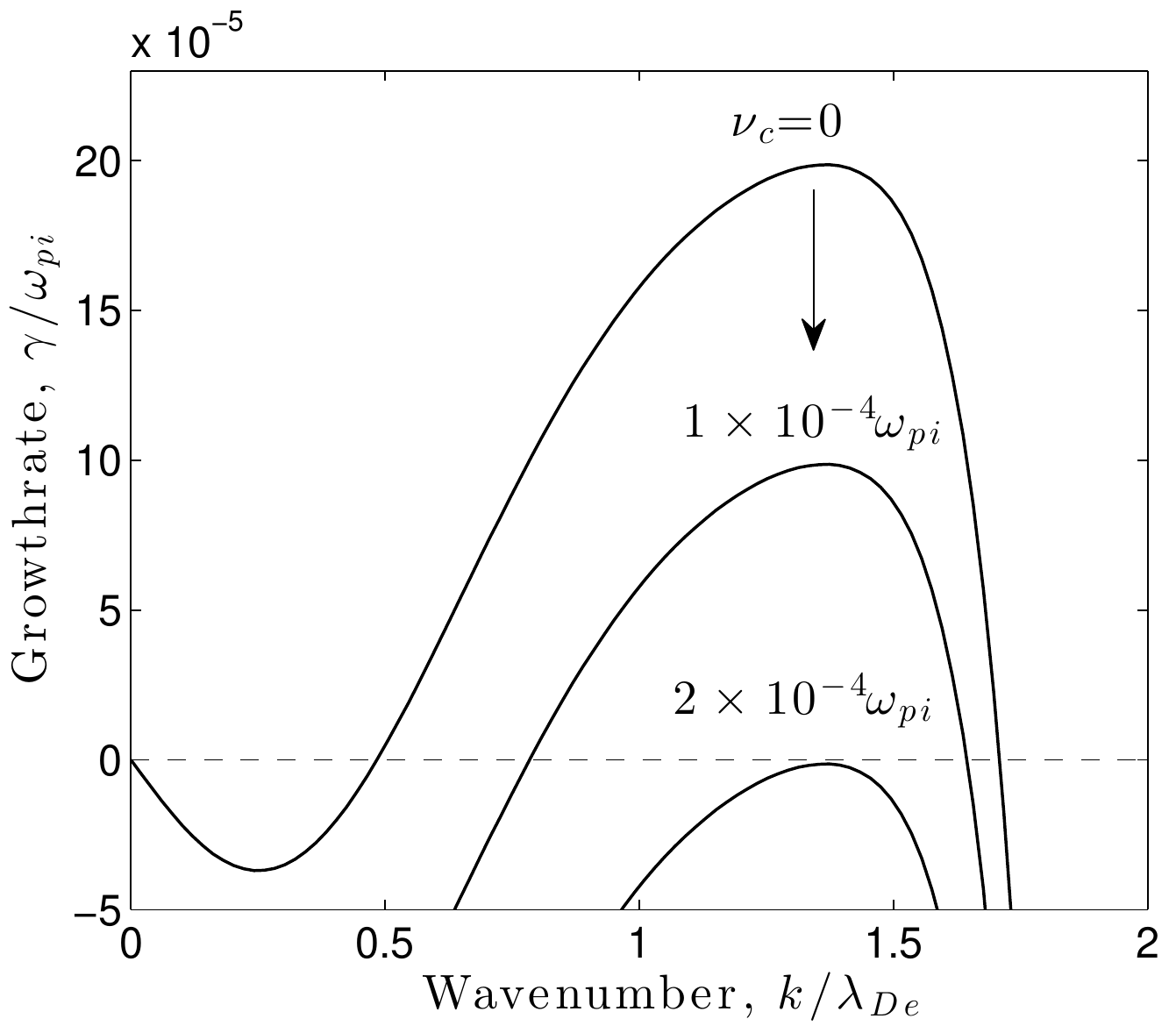}
\caption{Growth rate of a ion-acoustic instability calculated from equation~(\ref{eq:gama}) at the conditions: Ar$^+$, $T_e/T_i =100$, and $V_i = 0.9 c_s$. Curves for three model collision frequencies are shown: $\nu/\omega_{pi} = 0$, $1\times 10^{-4}$, and $2\times 10^{-4}$.  }
\label{fg:ia_n}
\end{center}
\end{figure}

A BGK-type collision model with a constant collision frequency modifies the dispersion relation simply by reducing the growth rate in equation~(\ref{eq:gama}) by the collision frequency $\nu$: $\gamma \rightarrow \gamma - \nu$. The affect of this on the growth rate is illustrated in figure~\ref{fg:ia_n}. A pressure stability diagram can then be constructed from the requirement that $\gamma_{\max} > \nu$, where $\gamma_{\max}$ is the growth rate based on the collisionless equation~(\ref{eq:gama}). Applying the above expression for the collision frequency in terms of the cross section along with the ideal gas equation of state at room temperature $n_g$ [cm$^{-3}$] = $3\times 10^{13} p$~[mTorr] provides an estimate for the neutral pressure conditions required for ion-acoustic instability 
\begin{equation}
p_c [\scriptsize{\textrm{mTorr}}] \lesssim \frac{4.5\times 10^{-17}}{\sigma [\textrm{cm}^2]} \frac{\gamma_{\max}}{\omega_{pi}} \frac{c_s}{V_i} \sqrt{\frac{n_e [\textrm{cm}^{-3}]}{T_e [\textrm{eV}]}} . \label{eq:iapressure}
\end{equation}

\begin{figure}
\begin{center}
\includegraphics[width=7cm]{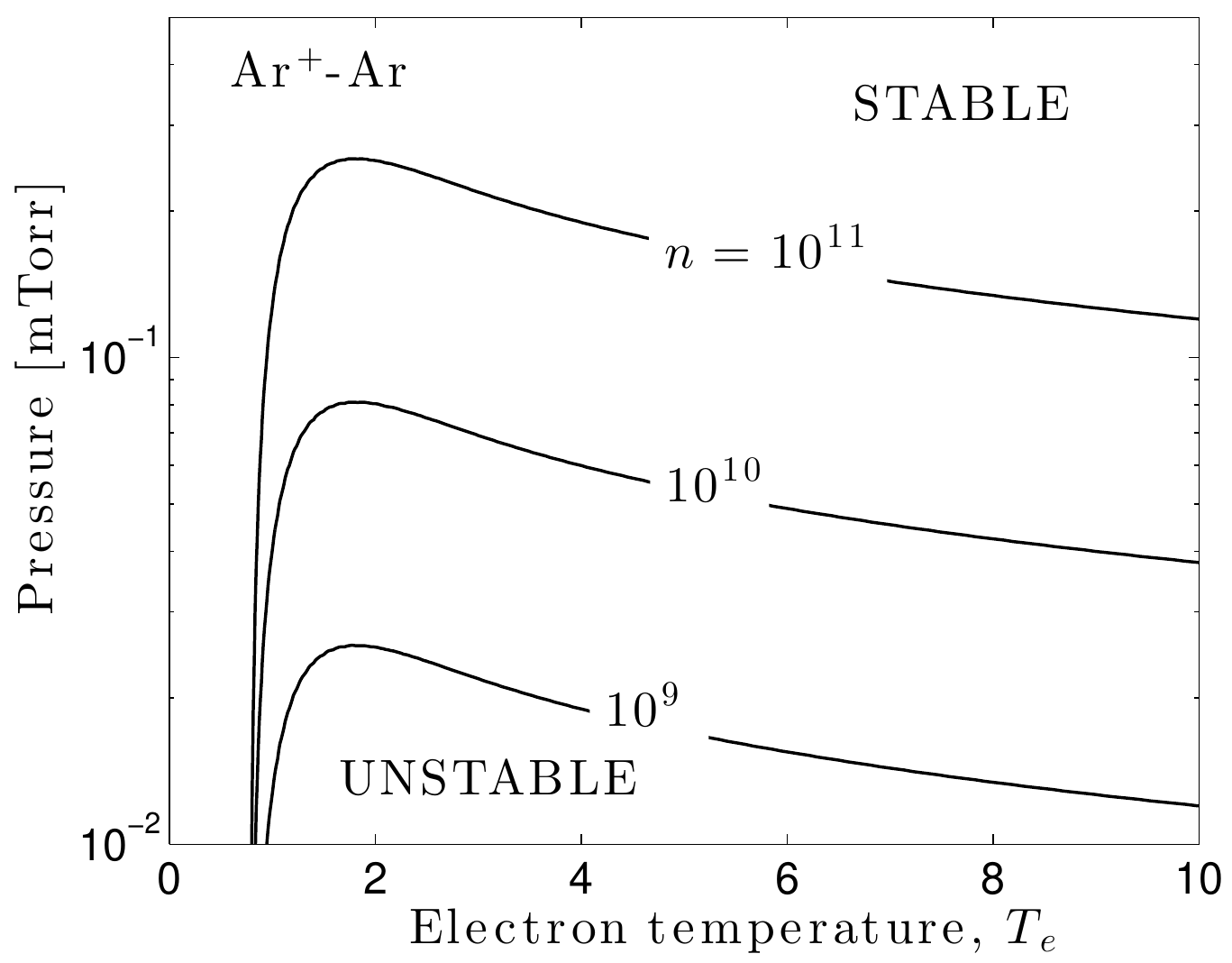}
\caption{Neutral gas pressure threshold for ion-acoustic instabilities in the presheath of an Ar$^+$ plasma computed from equation~(\ref{eq:iapressure}). Here, the parameters $V_i=c_s$, $T_i = 0.026$ eV, and three values of the electron density $10^9$, $10^{10}$ and $10^{11}$ cm$^{-3}$ have been chosen. }
\label{fg:ia_pressure}
\end{center}
\end{figure}

Figure~\ref{fg:ia_pressure} shows an evaluation of equation~(\ref{eq:iapressure}) for an Ar$^+$ plasma taking the ion temperature to be 0.026 eV and the ion flow velocity to be $c_s$, corresponding to a value expected at the sheath edge. The Ar$^+$-Ar collision cross section is taken to be  $\sigma \simeq 1 \times 10^{-14}$ cm$^{2}$, which is an approximate value of the total momentum transfer cross section from \cite{phel:94}. In general this is energy dependent, but for Ar$^+$-Ar collisions it is approximately constant over the energy range of interest ($1-10$ eV). The figure shows that neutral collisions have a significant affect on ion-acoustic instabilities in the presheath of an Ar$^+$ plasma. According to this collision model, they limit the region of potential instability to very low pressures, much less than a mTorr for typical discharge densities. Since the growth rate in units of $\omega_{pi}$ is approximately proportional to the square root of the mass ratio, $\gamma_{\max}/\omega_{pi} \sim \sqrt{m_e/m_i}$ at large temperature ratio $T_e/T_i \gg 1$ the pressure threshold values are expected to increase for lighter gases, and decrease for heavier gases. However, the actual values of these boundaries depend on the detailed ion-neutral cross sections. We also mention that this simple collision model effectively treats all collisions in the same statistical manner, with a constant collision frequency. A more realistic model should distinguish different collisions processes, particularly with regard to large angle, large momentum-transfer collisions. In particular, it is unclear if this gives an accrate description of charge-exchange collisions, which happen to be the dominant collision process for Ar$^+$-Ar at the conditions shown in figure~\ref{fg:ia_pressure}. Future work will be required to further asses the details of ion-neutral collisional damping. 

\subsection{Transport effects}

\subsubsection{Ion scattering} 

Ion-neutral collisions often influence the IVDF in a presheath. Typically this takes the form of a non-thermal tail extending from the flowing distribution to low energy~\cite{sher:01,clai:06}. Several theories have been developed, starting with the seminal work of Tonks and Langmuir~\cite{tonk:29}, to quantify the IVDF in this region. Accurate models are important for many applications. This is especially so in  materials processing because it influences the energy distribution of ions striking a material. 

It was recently predicted that if ion-acoustic instabilities arise in a presheath, wave-particle scattering can influence the IVDF~\cite{baal:11b}. In particular, the excitation of these weakly growing instabilities from the natural thermal fluctuations leads to an effective enhancement of the ion-ion Coulomb collision rate~\cite{baal:10}. This led to the prediction of a three-stage presheath with regard to the behavior of the IVDF~\cite{baal:11b}: (1) at the entrance, the IVDF is the bulk plasma distribution, which is often a Maxwellian, (2) as ions stream through the presheath and collide with neutrals or are produced by ionization, a non-thermal distribution forms that is flow-shifted and also has a tail extending to low energies, (3) if there is a region of ion-acoustic instabilities close to the sheath edge, the enhanced scattering causes the IVDF to thermalize to a flow-shifted Maxwellian near the sheath edge. The prediction that the wave-particle scattering thermalizes the distribution relies on the assumption that the amplitude of the fluctuations remains low enough that the dielectric response of the plasma is linear. If this case, one can consider the interaction between charged particles to occur via a dielectrically ``dressed'' Coulomb potential that includes the growing modes. The result is an enhanced collision rate, but other properties of standard Coulomb collisions, such as a unique Maxwellian equilibrium, are expected to hold~\cite{baal:10}. 

Recently, an experimental test of the three-stage presheath prediction was carried out using laser-induced fluorescence (LIF) in a xenon plasma~\cite{yip:15}. Measurements were consistent with the prediction at sufficiently low neutral pressure. By varying the neutral pressure and electron temperature, the pressure-temperature threshold for the thermalization near the sheath edge was measured using a criterion for the degree to which the IVDF can be considered Maxwellian. This was compared to a theoretical prediction based on an equation similar to equation~(\ref{eq:iapressure}). The experiment was consistent with the prediction and clearly measured the predicted $1/\sqrt{T_e}$ scaling. 

Kinetic Vlasov simulations have also been carried out to analyze the IVDF in the sheath and presheath~\cite{coul:15}. This work showed that collisionless bunching associated with the acceleration of ions through a sheath and presheath also causes the IVDF to become more localized and therefore appear thermalized. Future work is needed to understand, and distinguish, the combined effects of collisionless bunching and instability-enhanced collisions near the sheath edge. 


\subsubsection{Electron scattering} 

Since the ion sheath is very thin compared to the electron collision length scale in the vast majority of plasmas, it is nominally expected that the electron velocity distribution function (EVDF) near the sheath is severely depleted in the velocity phase-space region corresponding to electrons that traverse the ion sheath and escape to the boundary. However, some of the first measurements ever conducted in plasmas revealed that this region of velocity phase-space contained vastly more electrons than could be expected on theoretical grounds based on Coulomb collisions~\cite{lang:25}. This measurement of anomalous electron scattering has come to be known as Langmuir's paradox~\cite{gabo:55}. 

It has been proposed that ion-acoustic instabilities can enhance electron scattering in the presheath region~\cite{baal:09a}. The same instability-enhanced collision operator discussed in the context of ion-ion scattering in the previous section~\cite{baal:10} was also evaluated for electron-electron scattering the presence of ion-acoustic instabilities. The maximum collision rate enhancement in this kinetic theory occurs at velocities that are resonant with the phase speed of the wave. For electrons, the ion-acoustic speed is much less than the electron thermal speed, so the resonant region of phase-space is near the bulk of the distribution. The non-thermal region of interest near a sheath is in the tail of the distribution, which is far from this resonance. Nevertheless, the theory predicts that if the instabilities grow sufficiently in the presheath, they can significantly enhance scattering reaching even into the tail. The collision frequency scales with velocity approximately as $1/v^3$ from the resonant location. This is the same scaling that standard Coulomb collisions have referenced from the zero of the relative velocity vector $|\vc{v} - \vc{v}^\prime|$ (which is also near the bulk of the distribution). Thus, the instabilities enhance the Coulomb collision rate, but do not substantially change how it scales with velocity.   

The ion-acoustic instability boundaries discussed in the previous section determine when this affect might be expected to arise in low-temperature plasmas. Unfortunately, there is a dearth of experimental, or simulation, data to compare with the theory. A recent experiment by Godyak measured signficant depletion in the tail of the electron energy distribution function in argon discharges in the mTorr neutral pressure range~\cite{gody:15}. Figure~\ref{fg:ia_pressure} suggests that ion-acoustic instabilities would be damped in the presheath for this pressure. There is a great need for more experiments and simulations to understand electron scattering near sheaths. The stability boundaries provided here may contribute to determining when ion-acoustic instabilities can be expected to affect transport in this region.  


\section{Two ion species \label{sec:2ions}} 

\subsection{Linear waves}

\subsubsection{Ion-acoustic vs ion-ion two-stream}

If the plasma contains a mixture of two species of positively charged ions, the presheath electric field will generate a differential flow between ion species in addition to the differential flow between ions and electrons. Species can be distinguished by charge or mass. If the differential flow exceeds a threshold $\Delta V \equiv |V_1 - V_2| > \Delta V_c$, ion-ion two-stream instabilities can be excited. Depending on the values of the flows, and the differential flow, ion-ion two-stream instabilities can arise independently, or in addition to, ion-acoustic instabilities. When it does arise, the ion-ion instability often has a much larger growth rate than an ion-acoustic instability. 

\begin{figure}
\begin{center}
\includegraphics[width=7cm]{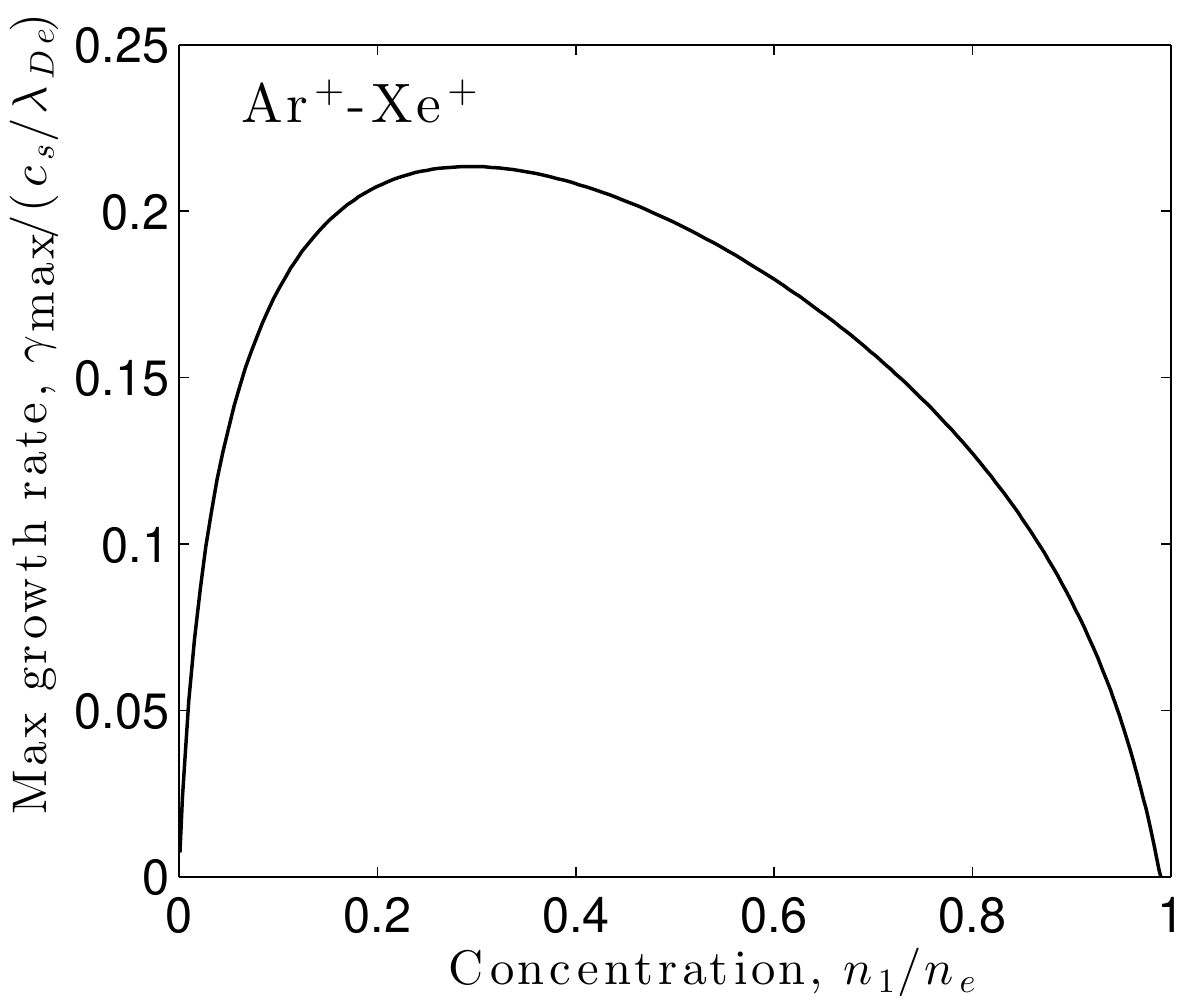}
\caption{Maximum growth rate of ion-ion two-stream instabilities for an Ar$^+$-Xe$^+$ plasma. The temperatures where chosen to be $T_e = 2$ eV, $T_i = 0.026$ eV, and the flow speed as $\Delta V = | c_{s1} - c_{s2}|$.}
\label{fg:gmax_c}
\end{center}
\end{figure}

To analyze the stability boundary of ion-ion two-stream instabilities,  we consider the standard linear dielectric response function for ion-frequency fluctuations for Maxwellian distributions 
\begin{equation}
\hat{\varepsilon} = 1 + \frac{1}{k^2 \lambda_{De}^2} \biggl[ 1 - \frac{z_1^2}{2} \frac{T_e}{T_1} \frac{n_1}{n_e} Z^\prime (\xi_1) - \frac{z_2^2}{2} \frac{T_e}{T_2} \frac{n_2}{n_e} Z^\prime (\xi_2) \biggr]  \label{eq:ephat}
\end{equation}
where $\xi_1 = \hat{k} \cdot \Delta \vc{V} (\Omega -1/2)/v_{T1}$, $\xi_2 = \hat{k} \cdot \Delta \vc{V} (\Omega + 1/2)/v_{T2}$, and $z_i$ is the ionic charge. Here, $Z$ is the plasma dispersion function and $Z^\prime (\xi) = dZ/d\xi$. The parameter $\Omega$ has been defined by the substitution 
\begin{equation}
\omega = \frac{1}{2} \vc{k} \cdot (\vc{V}_1 + \vc{V}_2) + \vc{k} \cdot \Delta \vc{V} \Omega , \label{eq:osub}
\end{equation}
where $\omega$ is the complex angular wave frequency. The ion-acoustic instability is excluded from equation~(\ref{eq:ephat}) because electrons are treated in the adiabatic limit. Thus, the ion-ion differential flow $\Delta \vc{V}$ arises in the growth rate calculation, but the ion-electron differential flow does not. 

Figure~\ref{fg:gmax_c} shows the maximum growth rate for ion-ion two-stream instabilities at the sheath edge of an Ar$^+$-Xe$^+$ plasma. This was calculated by solving equation~(\ref{eq:ephat}) numerically for $\Omega(k)$, then finding the peak of $\gamma (k)$. Common low-temperature plasma parameters have been chosen: $T_e=2$ eV, $T_i = 0.026$ eV, and $\Delta V = |c_{s1} - c_{s2}|$. Here, $c_{si} = \sqrt{k_B T_e/m_i}$ is the individual sound speed associated with species $i$, and $c_{s} \equiv \sqrt{(n_1/n_e) c_{s1}^2 + (n_2/n_e) c_{s2}^2}$ is a concentration-weighted system sound speed. Comparing this with figure~\ref{fg:ia_n} shows that the growth rate of the ion-ion instability in the mixture is approximately 10$^3$ times larger than the growth rate of ion-acoustic instabilities in a pure Ar$^+$ plasma at similar temperatures and flow speeds. Nominally, the ion-ion instability is larger by $\sim \sqrt{m_i/m_e}$. Thus, when the ion-ion instability is present it dominates over the ion-acoustic instability. 

\subsubsection{Two-stream: Different masses} 

A parameter space defining the ion-ion two-stream stability boundaries can be obtained directly by solving equation~(\ref{eq:ephat}) for the growth rate $\gamma (k)$, then determining the conditions at which $\gamma_{\max}=0$. However, the analysis can be simplified by noticing that the maximum growth rate typically occurs at a smaller wavenumber as the threshold is approached~\cite{baal:15}. Thus, the threshold boundaries can often be approximated from the long wavelength limit by taking $k=0$ in equation~(\ref{eq:ephat}). The threshold condition is then determined from
\begin{equation}
f(\Delta V_c) = z_1^2 c Z^\prime (\xi_1) + z_2^2 (1-c) Z^\prime (\xi_2) - 2 (T_i/T_e) = 0 .  \label{eq:f}
\end{equation}
where $c=n_1/n_e$ is the concentration of species 1. 

\begin{figure}
\begin{center}
\includegraphics[width=7cm]{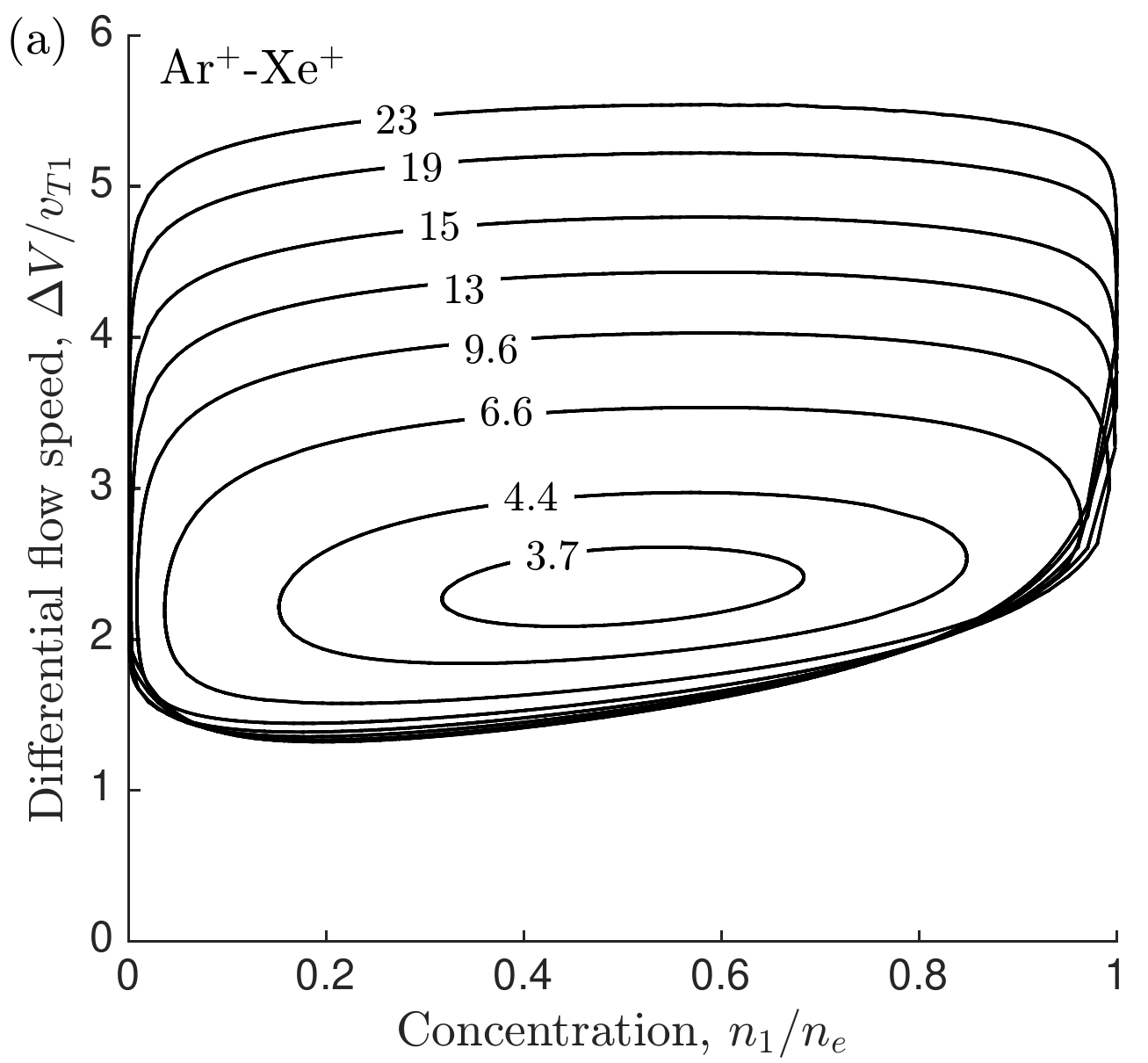}\\
\includegraphics[width=7cm]{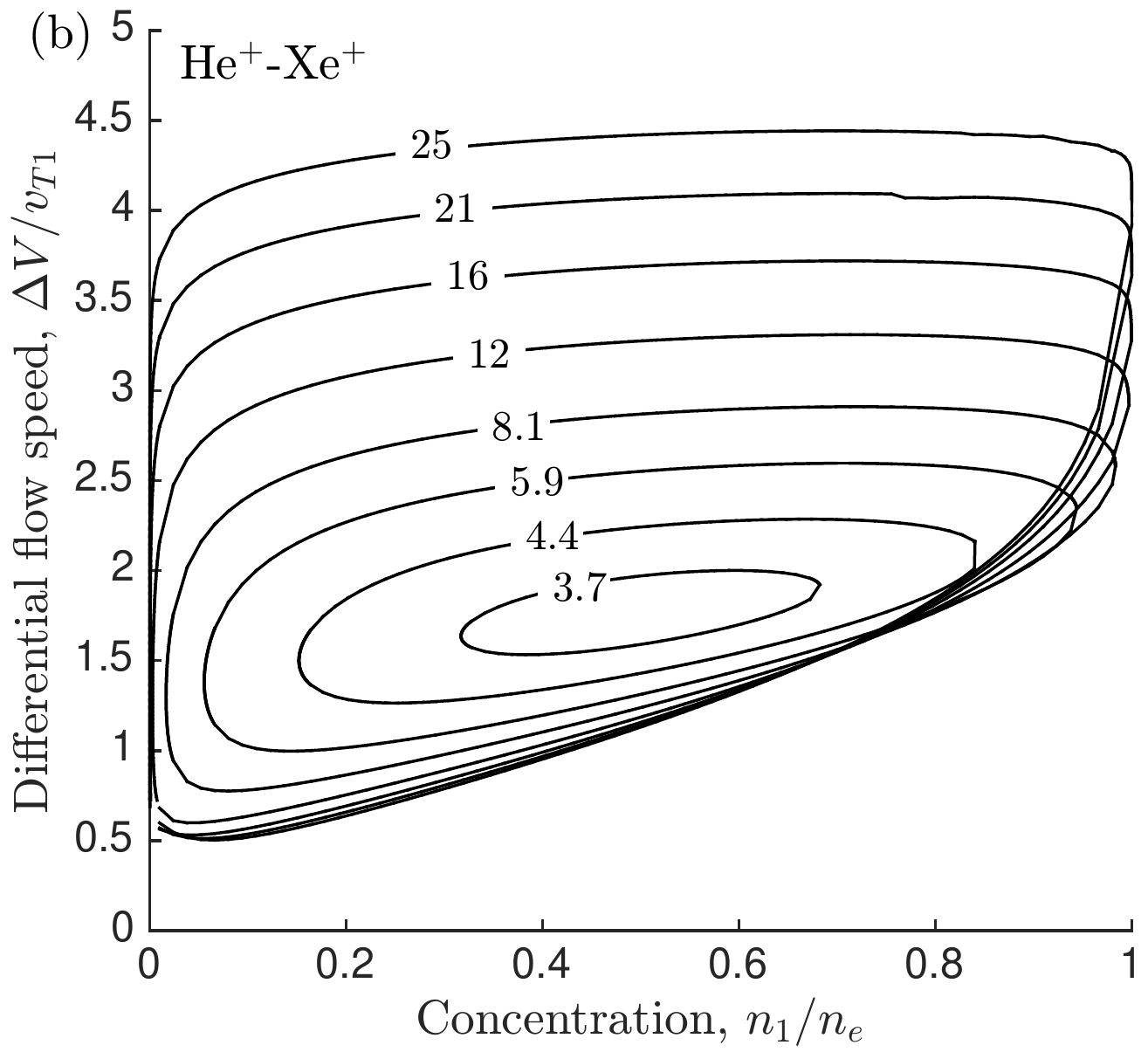}
\caption{(a) Two-stream instability boundaries in an Ar$^+$-Xe$^+$ plasma at several values of the temperature ratio $T_e/T_i$. Regions inside the closed contours are regions of predicted instability, while outside are regions of stability. (b) Analogous stability boundaries for He$^+$-Xe$^+$ plasma. }
\label{fg:dvc_wT}
\end{center}
\end{figure}

A recent analysis~\cite{baal:15} has explored the stability boundaries of binary ionic mixtures across a broad range of conditions using equation~(\ref{eq:f}). This includes determining the critical flow difference $\Delta V_c$ as concentration and $T_e/T_i$ are varied. It also includes determining a $T_e/T_i$ threshold for instability in a presheath by setting $\Delta V = |c_{s1}- c_{s2}|$. Figure~\ref{fg:dvc_wT} provides a further quantification of the stability boundaries for Ar$^+$-Xe$^+$ and He$^+$-Xe$^+$ mixtures, which are common in experiments~\cite{yip:10,hers:11}. This shows the stability boundary for the critical differential flow $\Delta V_c$ at several values of $T_e/T_i$. The region inside the closed loops is the unstable region of parameter space, while outside is stable. 



\subsubsection{Different charge states} 

\begin{figure}
\begin{center}
\includegraphics[width=7cm]{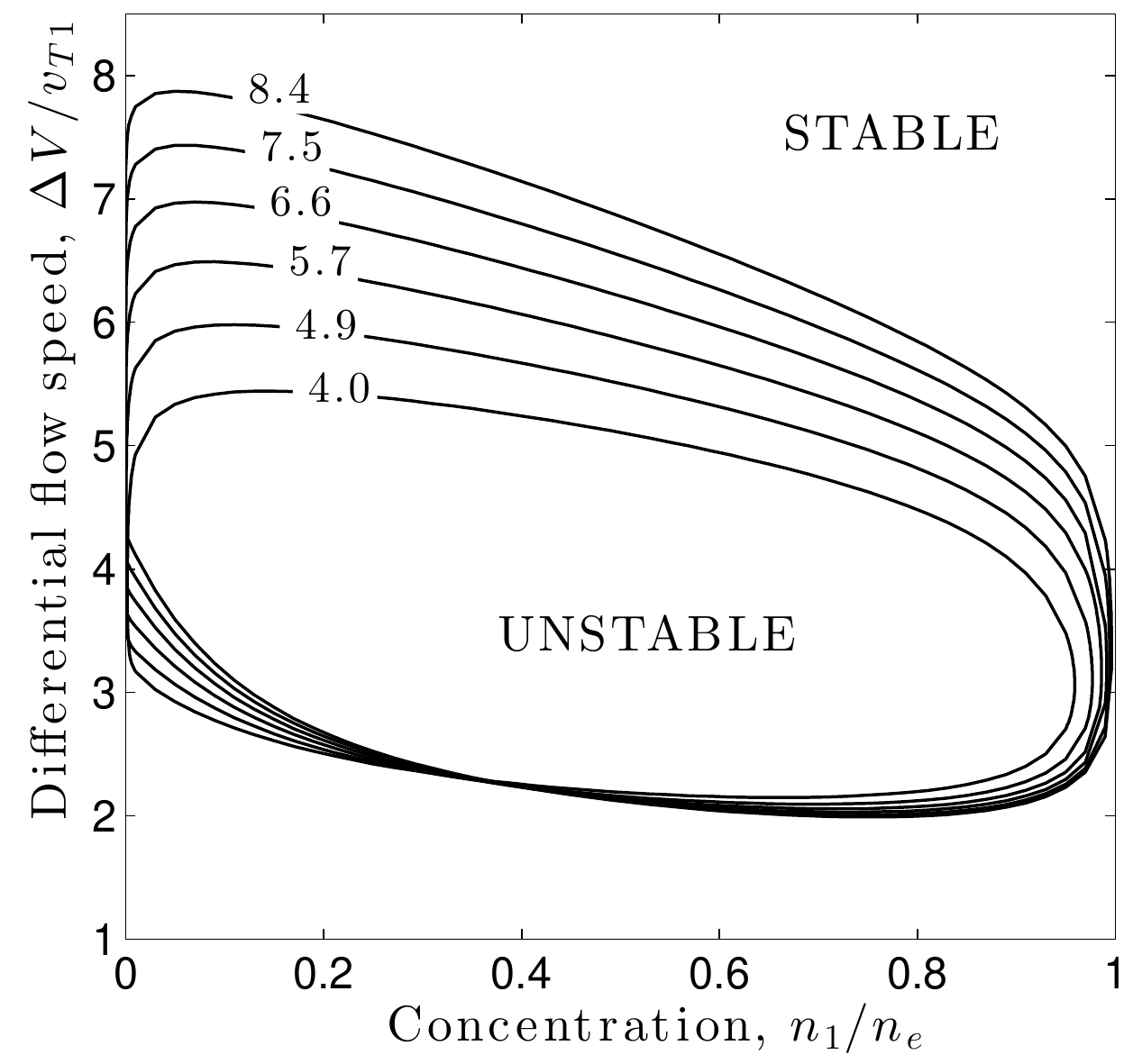}
\caption{Two stream stability boundaries for a binary ionic mixture with equal masses and charge states $z_1 = 1$, $z_2=2$ computed from equation~(\ref{eq:f}). Labeled curves correspond to different temperature ratios $T_e/T_i$.}
\label{fg:z1z2}
\end{center}
\end{figure}

Species can be distinguished not only by mass, but also by charge state. Figure~\ref{fg:z1z2} shows a stability diagram for a binary ionic mixture with equal masses, but charge states $z_1=1$ and $z_2=2$. Curves are shown at various temperature ratios $T_e/T_i$. In this type of a mixture, differential flow can be established in the presheath as the electric field accelerates the species with a higher charge to a faster speed. If ion-ion drag can be neglected, this leads to the expectation that the differential flow at the sheath edge is $\Delta V \simeq \sqrt{|z_1-z_2|k_BT_e/m_i}$. In most low-temperature plasmas, if electrons are hot enough to doubly ionize a fraction of atoms, then the concentration of the singly ionized species often remains much larger than the doubly ionized species (corresponding to the $n_1/n_e \ll 1$ region of figure~\ref{fg:z1z2}). 

\subsubsection{Neutral damping} 

\begin{figure}
\begin{center}
\includegraphics[width=7cm]{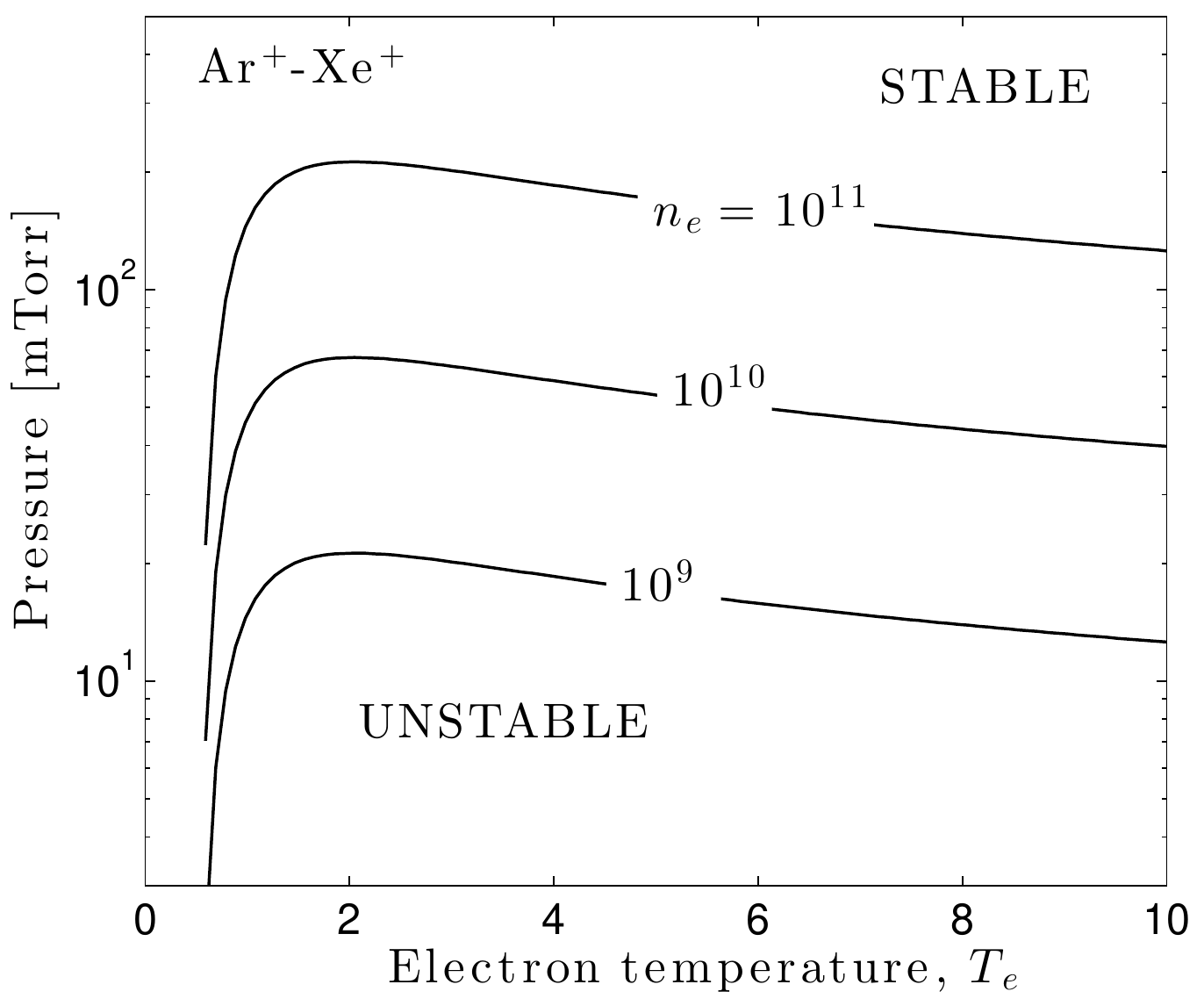}
\caption{Estimated neutral pressure threshold for ion-ion two-stream instabilities in an Ar$^+$-Xe$^+$ mixture. Here, $T_i = 0.023$ eV and $n_1=n_2$ have been chosen. }
\label{fg:2s_pressure}
\end{center}
\end{figure}

Since the growth rate of ion-ion two-stream instabilities typically far exceeds that of ion-acoustic instabilities in the presheath of binary ionic mixtures, the neutral pressure thresholds are also expected to be much higher. Figure~\ref{fg:2s_pressure} shows an estimate of the neutral pressure threshold in an Ar$^+$-Xe$^+$ mixture at different plasma densities for $n_1=n_2$ and $T_i =0.023$ eV. This was computed using equation~(\ref{eq:iapressure}) taking $\gamma_{\max} c_s/(\omega_{pi} V_i) \simeq \gamma_{\max}/(c_s/\lambda_{De})$, where $\gamma_{\max}$ was calculated from equation~(\ref{eq:ephat}). The ion-neutral collision cross sections for both Ar$^+$-Ar and Xe$^+$-Xe were approximated as $\sigma \simeq 1 \times 10^{-14}$ cm$^{2}$. This is supported by the total momentum scattering cross section data compiled in \cite{phel:94} and \cite{pisc:03}. Ar$^+$-Xe and Xe$^+$-Ar collisions were not included in this estimate. As expected, the figure shows much higher threshold values for the neutral pressure than were found for ion-acoustic instabilities in figure~\ref{fg:ia_pressure}. Since a neutral pressure of several mTorr is common in low-temperature plasma experiments, these figures suggest that in binary mixtures ion-ion two-stream instabilities may be a more common occurrence in the presheath than ion-acoustic instabilities.  

\subsection{Transport effects} 

When ion-ion two-stream instabilities arise in a presheath, it has been predicted that the ion-ion friction force is rapidly enhanced beyond the nominal Coulomb collision level. Since this occurs on a length scale that is much shorter than the presheath length scale, it led to the prediction that the differential ion flow speed cannot significantly exceed the threshold condition $\Delta V \leq \Delta V_c$ for instability onset. At the sheath edge, the generalization of Bohm's criterion to multiple ion species provides the following constraint on the ion speeds at the sheath edge~\cite{cook:80,riem:95}
\begin{equation}
\frac{n_1}{n_e} \frac{c_{s1}^2}{V_1^2} + \frac{n_2}{n_e} \frac{c_{s2}^2}{V_2^2} \leq 1 . \label{eq:bohm2}
\end{equation}
As in the single species case, equality is expected to hold in equation~(\ref{eq:bohm2}) in quiescent weakly-collisional plasmas~\cite{fran:00}. This has been confirmed experimentally~\cite{hers:11}. If ions are not collisionally coupled, it is nominally expected that each ion species obtains the same kinetic energy in response to falling through the presheath potential drop, $\frac{1}{2}m_1 V_1^2 = \frac{1}{2} m_2 V_2^2$. In this case, the solution of equation~(\ref{eq:bohm2}) is that each ion species obtains the individual sound speed at the sheath edge $V_1 = c_{s1}$ and $V_2 = c_{s2}$~\cite{lieb:05,fran:00}. However, if  $\Delta V_c < |c_{s1} - c_{s2}|$ two-stream instabilities are expected to arise in the presheath. In this case, the differential flow speed is expected to be $\Delta V_c$. Thus, a condition 
\begin{equation}
|V_1 - V_2| = \min \lbrace \Delta V_c, |c_{s1} - c_{s2}| \rbrace \label{eq:dvc}
\end{equation}
was suggested as a criterion that, along with equation~(\ref{eq:bohm2}), determines the speed of each ion species at the sheath edge~\cite{baal:11,baal:15}. 

The combination of equations~(\ref{eq:bohm2}) and (\ref{eq:dvc}) led to the prediction that there is a significant concentration dependence on the speed of ions at the sheath edge. At conditions of $T_e/T_i \gg 1$, the speed of each species is predicted to approach the common system sound speed $c_s$ when $n_1 \simeq n_2$, and trend toward the individual sound speeds ($c_{si}$) for dilute mixtures~\cite{baal:09,baal:11,baal:15}. For mixtures with large mass ratios, this can lead to a significant variation in the ion speed at the sheath edge as the concentration varies. For example, in He$^+$-Xe$^+$ at typically experimental conditions the He$^+$ speed changes by a factor of 2.

Several detailed tests of the predicted speeds have been conducted experimentally using laser-induced fluorescence including Ar-Xe discharges and He-Xe discharges~\cite{yip:10,hers:11}. These have shown excellent agreement with the theoretically predicted ion speeds, and have shown fluctuation measurements consistent with ion-ion two-stream instabilities~\cite{hers:05}. Recent PIC simulations have also shown similar agreement with the predicted ion speeds, and have also provided a detailed analysis of both the fluctuation spectrum near the sheath edge and the profile of friction in comparison to other terms of the ion momentum balance equation~\cite{baal:15}. Both of these also agree with the expectations of the theory. Other PIC simulations in mixtures have been conducted at conditions where two-stream instabilities are not predicted to occur in the presheath~\cite{gudm:11,levk:14} (see \cite{baal:15} for an analysis of the stability conditions). In both of these cases, the ion speeds were observed to be the individual sound speed at the sheath edge, which is also consistent with the theory. These experiments and simulations were conducted at low pressure, well below the thresholds indicated in figure~\ref{fg:2s_pressure} (with the exception of \cite{levk:14}, which was conducted at higher pressure, but this was also expected to be in a stable regime based on the temperature ratio in the simulation). No study has yet been conducted to explicitly test the pressure stability boundary.

\section{Discussion\label{sec:conc}} 

Standard linear electrostatic stability theory, which has been validated in many contexts, predicts that the ion flow through a presheath can excite instabilities under the conditions of low neutral pressure and large electron-to-ion temperature ratio $T_e/T_i \gg 1$. Although much has been done to quantify the stability boundaries for conditions relevant to low temperature plasmas, and several of the predicted transport effects have been tested experimentally, much remains to be done. This includes theoretical development to further understand the parameter regimes at which instability can be expected, as well as how the instabilities affect transport and how the change in transport might influence applications. Further experimental and simulation tests are required to validate these predictions. 

For single ion species plasmas, the major research need is a measurement or simulation showing the presence of ion-acoustic instabilities in parameter regime predicted to be unstable. There is currently no such direct measurement. Only the indirect effect of the predicted ion thermalization has been measured~\cite{yip:15}. On the theoretical side, further developments are needed to understand how ion-neutral collisions damp the instability. A simple model based on a constant collision frequency was provided here, but it is unclear how accurate this is, especially for charge-exchange collisions. This may significantly influence the stability boundaries. Further analysis is also required to validate the predicted enhancement of electron scattering. 

For binary ionic mixtures, much analysis and experimental validation has already been completed. A primary research need going forward is a detailed analysis and validation of the neutral pressure threshold. This is important since discharges span a broad range of neutral pressures, extending across the estimated boundaries shown in figure~\ref{fg:2s_pressure}. It will also be important to establish how the predicted change in ion flow speed at the sheath edge affects plasma modeling in different situations, such as global models~\cite{kim:15} or models of plasma-materials interactions. Although the speed predicted by the instability-enhanced friction theory can differ substantially from the traditional individual sound speed prediction, it is still unknown if this has significant practical implications. 




\section*{Acknowledgments}

The author gratefully acknowledges Patrick Adrian for assistance in creating figures 8-10.
This work was supported by the Office of Fusion Energy Sciences at the U.S. Department of Energy under contract DE-AC04-94SL85000.



\end{document}